\documentclass[10pt]{iopart}

\usepackage{iopams}
  \expandafter\let\csname equation*\endcsname\relax
  \expandafter\let\csname endequation*\endcsname\relax 

\usepackage{amsfonts,bm,amsmath}

\usepackage{graphicx}

\newcommand*{\onlinecite}[1]{\cite{#1}}

\begin{document}
\title[Efficiency of the biexciton admixture mechanism for
MEG]{Efficiency of the coherent biexciton admixture mechanism for
  multiple   exciton generation in InAs nanocrystals}  
\author{Piotr Kowalski, Pawe{\l} Machnikowski}
\address{Department of Theoretical Physics, Wroc{\l}aw University of Technology, 
50-370 Wroc{\l}aw, Poland}
\ead{\mailto{pawel.machnikowski@pwr.edu.pl}}

\begin{abstract}
We study the coherent mixing between two-particle (single exciton) and
four-particle (biexciton) states of a semiconductor nanocrystal
resulting from the coulomb coupling between states with different
numbers of electron-hole pairs. Using a simple model of the
nanocrystal wave functions and an envelope function approach, we
estimate the efficiency of the multiple exciton generation (MEG) process
resulting from such coherent admixture mechanism, including all the
relevant states in a very broad energy interval. We show that in a
typical ensemble of nanocrystals with 3~nm average radius, the
onset of the MEG process appears about 1~eV above the lower edge of
the biexciton density of states and the efficiency of the process
reaches 50\% for photon energies around 5~eV. The MEG onset shifts to
lower energies and the efficiency increases as the radius grows. We
point out that the energy dependence of the MEG efficiency differs
considerably between ensembles with small and large inhomogeneity of
nancrystal sizes.
 \end{abstract}


\maketitle

\section{Introduction}\label{sec:introduction}

Confining the carriers to a nanometer-scale volume of a
semiconductor nanostructure not only leads to qualitative
modification of the energy spectrum but also changes the relative role
of various kinetic processes that take place in these systems. In
particular, it has been proposed \cite{nozik02} that an enhancement of
impact ionization processes in nanostructures can be exploited to
increase the efficiency of solar cells by enabling multiple exciton
generation (MEG) upon absorption of a single high-energy photon. In
spite of the initial controversy concerning the actual efficiency of
this process
\cite{schaller05,schaller06,schaller07,trinh12,nair07,ben-lulu08},
which arouse from the experimental difficulty in extracting the real
MEG yields \cite{pijpers07,mcguire10,binks11}, 
direct measurements of photocurrent from nanocrystal-based structures
\cite{sambur10,semonin11} confirm the initial expectation that
efficiency enhancement due to the MEG process is feasible.

In the single-particle picture, only the transitions leading to creation of a single
electron-hole pair (exciton) are optically active. Therefore, the MEG
process requires Coulomb interactions that couple configurations with
different numbers of excitons. Such inter-band Coulomb couplings are
indeed present in semiconductor nanocrystals and can be theoretically
computed by various methods 
\cite{deuk11,deuk12,franceschetti06,rabani08,califano09,%
baer12,allan06,delerue10,shabaev06,witzel10,silvestri10,korkusinski10}.
In their presence, there are various scenarios that can lead to the MEG effect. 
For instance, one can think of a sequential process in which the
initially created high-energy exciton subsequently decays into the
biexciton states that form a quasi-continuum with very high density of
states at sufficiently high energies. The rates for such a process have
been calculated using atomistic methods
\cite{rabani08,deuk11,deuk12}. On the other hand, biexciton states can
be created coherently due to the Coulomb-induced mixing between the
single-exciton (X) and biexciton (BX) states \cite{Kowalski:2013,Azizi2015}. In
this picture, admixture of BX states to X states
increases the average number of excitons in the resulting few-particle
eigenstate, while admixture of certain X states to BX states
can make the latter optically active. As a result, the nanocrystal
state emerging from a single photon absorption can involve more than
one electron-hole pair on the average.

In this paper, we estimate the efficiency of MEG resulting from the
admixture mechanism in an InAs nanocrystal. We use the method for
calculating the Coulomb 
couplings between X and BX states based on the envelope function
(\mbox{$\bm{k}\cdot\bm{p}$}) approach developed in our previous paper
\cite{Kowalski:2013}, which is a low-cost approximate method that allows us to
build statistics over very many X and BX states in a broad energy
window and for a distribution of nanocrystal sizes. The predicted
efficiencies are in the range of several up to 50\% for photon
energies below 5~eV, which is roughly consistent with experimental findings
\cite{nair08,mcguire10}. 

The paper is organized as follows. In Sec.~\ref{sec:model}, we
present the model and the method used to estimate the
efficiencies. The results are discussed in
Sec.~\ref{sec:results}. Sec.~\ref{sec:summary} concludes the paper.

\section{Model and method}\label{sec:model}

We use the simple model of a spherical nanocrystal as in 
Ref.~\onlinecite{Kowalski:2013}. The carrier states are modeled by
single-band wave functions with envelopes corresponding to a simple
spherical potential well with infinite barriers, with a constant hole
effective mass and a self-consistent, energy-dependent electron mass
\cite{efros98,Kowalski:2013}. Thus, each single particle state is
characterized by the band and the three quantum numbers $(n,l,m)$ for
the envelope wave function. Two kinds of few-particle configurations
are relevant here: 
The first group are optically active (bight) states with one
electron-hole (e-h) pair, referred to as single 
exciton (X) states. In our model, for such states to be bright (in the
sense of a dipole-allowed transition), the
quantum numbers $n,l$ of the electron and the hole must be the same,
while the values of $m$ must be opposite. In addition, the band
angular momentum (``spin'') projection of the hole must be specified to
fully characterize the state. Only the states with holes from the
$j=3/2$ valence band can be bright. The second group are states
with two e-h pairs, referred to as biexciton (BX) states. These are
labeled by the full set of quantum numbers for the two electron and
two hole states, as well as the spin (singlet-triplet) configuration
of the two electrons and the subband (``spin'') configuration
of the two holes. 

Only the diagonal (first order)
Coulomb correction to the energies of X and BX states as well as
the electron and hole exchange energies for the BX states are
taken into account. This is done on the usual mesoscopic level of
envelope functions. 
In addition, the intraband Coulomb coupling (i.e.,
coupling between configurations with a different number of e-h pairs),
which is relevant for the effect to be discussed here, is
included. On the mesoscopic level, these Coulomb terms vanish in the
single band approximation due to orthogonality of Bloch functions
\cite{Kowalski:2013,Azizi2015}, hence they are taken on the
microscopic level involving the Bloch functions, in the first
order of the expansion in terms of $a/R$, where $a$ is the lattice
constant and $R$ is the nanocrystal radius (see
Refs.~\onlinecite{Kowalski:2013,Azizi2015} for details). All the
Coulomb couplings taken into account include both the direct
interaction term as well as interaction mediated by surface
polarization which is due to the dielectric discontinuity on the
nanocrystal border. Here we only include couplings corresponding to
creation of an e-h pair with electron intraband relaxation, with the
other hole being a ``spectator''. The values of all the parameters,
corresponding to an InAs nanocrystal, are as in
Ref.~\onlinecite{Kowalski:2013}.  

An
inhomogeneous ensemble of nanocrystals is modeled by assuming a
Gaussian distribution of nanocrystal radii $f(R)$, with the average
$R_{0}$ and standard deviation $\sigma_{R}$. In practice, the
size dependence of the Coulomb couplings is given by a simple $1/R$ or
$1/R^{2}$ scaling \cite{Kowalski:2013}, while the energies have been
computed for 7 values of $R$ in the range $(2.7,3.3)$~meV and
approximated by a quadratic fit, which yields a good quantitative
approximation in the required range of nanocrystal radii.

The BX admixture to an X state $|X_{j}\rangle$ in a nanocrystal of a
given radius is obtained (as in Ref.~\cite{Kowalski:2013})
by diagonalizing the Hamiltonian including only the X state in
question and the BX sates $|BX_{i}\rangle$ directly and sufficiently
strongly coupled to $|X_{j}\rangle$. The criterium for the selection of
these ``sufficiently strongly coupled'' BX states out of the infinite
number of such states is as follows: If $h_{ji}$ is the magnitude of the
coupling matrix element between the two states and $\Delta
E_{ji}$ is the energy separation between them, then a given
state BX is included if the ratio 
$q_{ji}=|h_{ji}/\Delta E_{ji}|$ is larger than
$0.01$. Note that $|q_{ji}|^{2}$ is the perturbative measure of the admixture
of a given state BX to X. After diagonalization of the Hamiltonian
constructed in this way, we select the eigenstate
$|\Psi_{0}^{(j)}\rangle$ with the highest X contribution as the new nominally
single-exciton state and its BX admixture is determined as 
\begin{displaymath}
x_{\mathrm{BX/X}}^{(j)}=
\sum_{i}\left|\left\langle BX_{i}|\Psi_{0}^{(j)} \right\rangle \right|^{2} =
1-\left|\left\langle X_{j}|\Psi_{0}^{(j)} \right\rangle \right|^{2}.
\end{displaymath}

In the calculation of the MEG efficiency as a function of the photon
energy, we need the relative absorption by a given state and the resulting
number of e-h pairs. For a nominally single-exciton state, the former
is proportional to the brightness $S_{j}=(1-x_{\mathrm{BX/X}}^{(j)})s_{j}$ of this
state, where the nominal brightness $s_{j}=1$ for a heavy-hole exciton and
$s_{j}=1/3$ for a light-hole exciton. The contribution of this state to
the overall absorption of an inhomogeneous ensemble is
$f(R)S_{j}$. On the other hand, its relative contribution to the e-h pair
production is $f(R)S_{j}(1+x_{\mathrm{BX/X}}^{(j)})$.

The X admixture to bright BX states is determined in a similar way. 
We diagonalize the Hamiltonian containing only the BX state in
question (denoted $|BX_{j}\rangle$) and all the bright X states
$|X_{i}\rangle$ directly and sufficiently 
strongly coupled to the BX state (those for which 
$q \geq  0.01$). We identify the eigenstate $|\Phi_{0}^{(j)}\rangle$ 
which has the biggest BX component and treat it as the new, corrected
BX state. The single-exciton admixture to this state is equal to
\begin{displaymath}
x_{\mathrm{X/BX}}^{(j)} =  
1 - \left| \left\langle BX_{j}\left| \Phi_{0}^{(j)}\right.\right\rangle \right|^{2}.
\end{displaymath}
The brightness of the BX states is 
$S_{j} = \sum_{i}x_{\mathrm{X/BX}}^{(ji)} s_{i}$, 
where $x_{ji}$ and $s_{i}$ are the admixtures of various X states and
their nominal brightness, respectively (in the vast
majority of cases, there is only one sufficiently strongly coupled X
state). The relative contribution to the absorption is $f(R)S_{j}$ 
and the relative contribution to the e-h pair production is
$f(R)S_{j}(2-x_{\mathrm{X/BX}}^{(j)})$. 

In practice, since the largest Coulomb couplings found for our model
of the nanocrystal with a 3~nm radius are around 40~meV, we include
admixed states with energies up to 10~eV (in the whole relevant range
of radii) in order to provide quantitatively correct results in the
5~eV energy range of interest.

We then simulate an inhomogeneous nanocrystal ensemble by sampling the
radius distribution at 0.003~nm intervals and compute the dependence
of the MEG efficiency on energy as a histogram with the energy axis divided
into finite bins of $\delta E=20$~meV width. In the energy bin $l$
with a central energy $\mathcal{E}_{l}$, the total
absorption rate is calculated (up to a constant factor) as 
\begin{displaymath}
\alpha_{l} = 
\sum_{nj}f(R_{n})S_{j}(R_{n})\Theta(\delta E/2 - |\mathcal{E}_{l}-E_{j}(R_{n})|),
\end{displaymath}
where $n$ numbers the nanocrystals in the simulated ensemble, $j$ runs
through all the X and BX states, $\Theta$ is the Heavyside step
function, and we have explicitly witten the $R$-dependence of the
energy and brightness of the state $j$.
The e-h pair production rate is 
\begin{align*}
\beta_{l} &= 
{\sum_{nj}}'f(R_{n}) S_{j}(R_{n}) (1+x_{\mathrm{BX/X}}^{(j)})\\
&\quad\times\Theta(\delta E/2 -
|\mathcal{E}_{l}-E_{j}(R_{n})|)\\
&\quad +{\sum_{nj}}''f(R_{n}) S_{j}(R_{n}) (2-x_{\mathrm{X/BX}}^{(j)}) \\
&\quad\times\Theta(\delta E/2 -
|\mathcal{E}_{l}-E_{j}(R_{n})|),
\end{align*}
Where the first sum runs through the X states and the second one
through the BX states.
The MEG efficiency is then 
\begin{displaymath}
F(\mathcal{E}_{l}) = \frac{\beta_{l}}{\alpha_{l}} -1,
\end{displaymath}
that is, $F=1$ corresponds to 100\% biexciton generation.
 
 \section{Results}\label{sec:results}

\begin{figure}[tb]
\begin{center}
 \includegraphics[width=8.5cm]{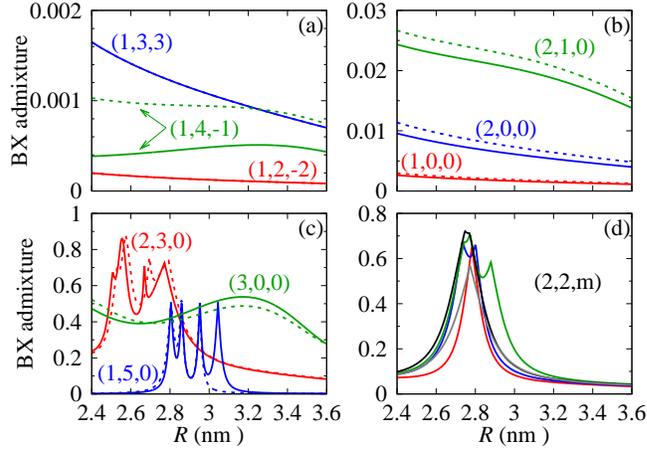}
\end{center}
 \caption{\label{fig:Xmixture}Admixture of BX states to bright X states as a
   function of the nanocrystal radius. The states are labeled by their
   quantum numbers $n,l,m$. Solid (dashed) lines correspond to
   excitons with the hole spin projections of $\pm 3/2$ and $\pm
   1/2$, respectively. In (a,b) selected low-energy
   states are shown. In (c,d) we show some states of higher energy.} 
\end{figure}

We begin the discussion of the results from an overview of the typical values
of X and BX admixtures to the BX and X states, respectively, as well
as their dependence on the nanocrystal radius. 

In Fig.~\ref{fig:Xmixture}(a,b) we show the BX admixture to a few selected
low-energy states (the exact information on the
energies is contained in the Appendix) as a function of the radius
$R$. The value of this admixture remains very low even for states up
to 1~eV above the lowest BX state, which is due to selection rules
that exclude coupling between excited bright X states and the lowest
BX states \cite{Kowalski:2013}. The admixture to X
states with $\pm 1/2$ hole spin projection is typically slightly
higher than to those with $\pm 3/2$ hole spin.

Fig.~\ref{fig:Xmixture}(c,d) presents the BX admixture for selected
states with higher energies. In general, much higher values are
obtained in this case. The $R$ dependence clearly has a resonant
character and peaks for some nanocrystal radii, when the coupled X and
BX states cross.
As shown in Fig. ~\ref{fig:Xmixture}(c,d), there is only very weak
dependence on the value of the quantum number $m$ and on the hole spin
projection in the X state. 

\begin{figure}[tb]
\begin{center}
 \includegraphics[width=6.8cm]{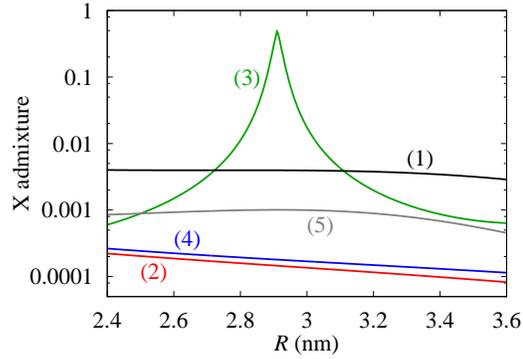}
\end{center}
\caption{\label{fig:BXmixture} X admixture to selected BX states.}
\end{figure} 

Fig.~\ref{fig:BXmixture} shows the admixture of X states to BX
states.  In most cases, this value does not exceed $0.01$,
irrespective of the energy of a given BX state (see Appendix for
detailed information on the states selected for this figure)
hence these states remain nearly completely dark. However, as
illustrated by the case (3) in
Fig.~\ref{fig:BXmixture}, resonant enhancement of the admixture may
happen in the case of a resonance between the chosen BX 
state and one of the X states. Then, in a relatively narrow 
peak, admixture reaches 0.5.  In spite of the large number of states
sufficiently strongly coupled   
to one of the bright states with energy below 5~eV (about 7500 BX
states for $R=3$~nm in our model), there are only several tens of such
resonances in this energy window. 

\begin{figure}[tb]
\begin{center}
 \includegraphics[width=85mm]{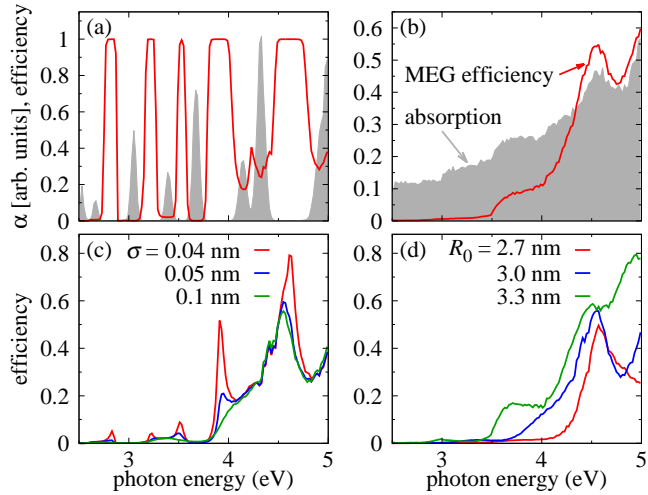}
\end{center}
 \caption{The MEG efficiency (lines) as a function of the photon energy: (a)
   for a very weakly inhomogeneous ensemble ($R_{0}=3$~nm,
   $\sigma=0.02$~nm), compared with the absorption 
   coefficient (shaded area); (b) as in (a) but for a much larger ensemble
   inhomogeneity ($\sigma=0.3$~nm); (c) as a function of the inhomogeneity for
   $R_{0}=3$~nm; (d) as a function of the average radius for $\sigma=0.15$~nm. }
 \label{fig:etaMEG}
\end{figure}

Finally, Fig.~\ref{fig:etaMEG} shows the dependence 
of the MEG efficiency on the photon energy. In a nearly homogeneous
ensemble of nanocrystals (Fig.~\ref{fig:etaMEG}(a)), the efficiency reaches
1 in certain energy intervals. Comparison with the absorption spectrum
shows that this happens only for energies where absorption is nearly
absent. Clearly, the reason for such an efficient biexciton generation
is the discrete nature of the single exciton density of states: in a
weakly inhomogeneous ensemble there are extended energy intervals
where no bright single exciton states are present so that all the
absorption originates from the quasi-continuous background of nearly
purely biexcitonic states.

This picture is different in an ensemble of nanocrystals with a
broader size distribution (Fig.~\ref{fig:etaMEG}(b)). Now both the
absorption and the MEG 
efficiency become quasi-smooth functions of the photon energy and both
grow nearly monotonically. The onset of the MEG (the MEG threshold) is
clearly marked at 3.5~eV, which should be contrasted with the nominal
onset of the BX density of states at $\sim 2.5$~eV for $R=3$~nm.

The evolution of the energy dependence of
the MEG efficiency with increasing inhomogeneity is shown in more
detail in Fig.~\ref{fig:etaMEG}(c). Here one can see that already at
$\sigma=0.04$~nm the overlap of the inhomogeneously broadened
absorption peak is sufficient to suppress the large values of the
efficiency, leafing only a few marked maxima. These MEG efficiency
peaks are then washed out as the inhomogeneity grows further. 

In Fig.~\ref{fig:etaMEG}(d) we show the MEG efficiency for three
ensembles with different average nanocrystal radii. The two features
that can be noticed are the a shift of the MEG
threshold towards lower energies and larger overall values of the MEG
efficiency for larger nanocrystals.
 
 \section{Conclusions}\label{sec:summary}

We have estimated the quantum efficiency of the multiple exciton
generation via coherent, Coulomb-induced mixing of bright exciton and
biexciton states. Although the mixing is, in general, rather weak due
to small values of the Coulomb coupling elements (not exceeding a few
tens of meV), it can become much stronger near the crossing point
between the X and BX energies at particular values of the nanocrystal
radius. The relatively low computational cost of the envelope function
method used in our calculations has allowed us to include all the
relevant states in a 10~eV energy window.

An interesting property that emerges from our calculations is the high
MEG yield in the energy intervals where the single exciton density of
states vanishes in a weakly inhomogeneous ensemble. Although this
result has been obtained in our simple model of nanocrystal wave
functions, it essentially follows from the discrete nature of
single-exciton spectrum for moderate energies and the much more dense,
quasi-continuous spectrum of biexciton states. Therefore, it should be
a general feature of highly homogeneous  nanocrystal
ensembles. Whether this can be exploited in 
applications depends on the technological feasibility of building a
structure in which the very weak absorption of biexciton states is
accumulated to yield a considerable overall carrier injection.

In contrast, for less homogeneous nanocrystal ensembles, both the
absorption and MEG efficiency are smooth functions of the photon
energy, with a threshold at about 3.5~eV for nanocrystals with 3~nm
radius, which is 1~eV above the formal onset of the biexciton density
of states. The threshold shifts to lower energies for larger
nanocrystals. The efficiencies of the MEG process reach 50\% for the
photon energies about 5~eV and average radius of 3.0~nm and increase
for larger nanocrystals.

\appendix
\section{Information on the states used in the figures}

In this appendix we present the detailed information about the states
shown in Fig.~\ref{fig:Xmixture} and Fig.~\ref{fig:BXmixture}.

\begin{table}[tb]
\begin{center}
\begin{tabular}{|cc|ccc|}
\hline
\multicolumn{2}{|c|}{state} & \multicolumn{3}{|c|}{energy (eV)} \\
\hline
$n$ & $l$ & $R=2.4$~nm & $R=3.0$~nm & $R=3.6$~nm \\
\hline
1 & 0 & 1.59 & 1.30 & 1.11 \\
1 & 1 & 2.34 & 1.86 & 1.57 \\
1 & 2 & 3.11 & 2.45 & 2.03 \\ 
1 & 3 & 3.92 & 3.05 & 2.51 \\
1 & 4 & 4.77 & 3.67 & 3.00 \\
1 & 5 & 5.66 & 4.32 & 3.52 \\
2 & 0 & 3.41 & 2.67 & 2.21 \\
2 & 1 & 4.39 & 3.39 & 2.78 \\
2 & 2 & 5.41 & 4.14 & 3.37 \\ 
2 & 3 & 6.47 & 4.91 & 3.97\\
3 & 0 & 5.66 & 4.31 & 3.50 \\
\hline
\end{tabular}
\end{center}
\caption{\label{tab-X}Energies of single exciton states for three
  values of the nanocrystal radius.}
\end{table} 

Table~\ref{tab-X} shows the energies of all the bright single exciton
states with energies below 5~eV at $R=3$~nm. The values are shown for
three nanocrystal radii. In our model, bright
states are those with identical quantum numbers $n,l$ for the electron and
the hole (shown in the leftmost columns). Shifts induced by
mixing with different BX states lead to differences between the energies of X states with
different values of $m$ as well as between states with different
projections of the hole spin. However, these differences are very
small, within 10~meV, hence we show only the energies of the state
with 3/2 hole spin projection and $m=0$.

\begin{table}[tb]
\begin{center}
\begin{tabular}{|c|cccccc|ccc|}
\hline
&\multicolumn{6}{|c|}{state} & \multicolumn{3}{|c|}{energy (eV)} \\
&e & e & h & h & $\Sigma_{e}$ & $\Sigma_{h}$ & 2.4~nm & 3.0~nm & 3.6~nm \\
\hline
(1) & 100 & 100 & 21$\overline{1}$ & 200 & $S$ & $T_{0}^{(3/2)}$
 & 4.66 & 3.56 & 2.92 \\
(2) & 111 & 100 & 210 & 110 & $T_{+}$ & $S_{\uparrow\Downarrow}$
 & 4.87 & 3.87 & 3.13 \\
(3) & 100 & 100 & 151 & 14$\overline{1}$ & $S$ & $S_{\uparrow\Downarrow}$
 & 5.78 & 4.31 & 3.47 \\
(4) & 210 & 100 & 21$\overline{1}$ & 110 & $S$ & $T_{+}^{(1/2)}$
 & 6.21 & 4.84 & 4.01 \\
(5) & 121 & 100 & 230 & 12$\overline{2}$ & $T_{0}$ & $S$
 & 6.55 & 4.96 & 4.02 \\
\hline
\end{tabular}
\end{center}
\caption{\label{tab-BX}Quantum numbers, spin configurations, and
  energies of biexciton  states used in
  Fig.~\ref{fig:BXmixture}. Labels (1)--(5) refer to that figure.} 
\end{table} 

Table \ref{tab-BX} shows the full data on the states presented in
Fig.~\ref{fig:BXmixture}. Here the first four columns contain the
quantum numbers $nlm$ for the two electrons and the two holes
(negative values of $m$ are denoted by a bar over the number) and the
next two columns show the spin configurations: For the electron, this
is just singlet ($S$) or one of the three triplet states  ($T_{0,\pm}$) with the
total spin projection 0 or $\pm 1$. For two-hole states with both
holes with $\pm 3/2$ or with both holes with $\pm 1/2$ spin
projection, the basis configurations are also singlet or triplet, with
the additional upper index $(3/2)$ or $(1/2)$, respectively. In
addition, configurations in which one hole has a 1/2 spin projection
and the other one has 3/2 spin projection appear in our
calculations. For these, we introduce the symmetrized and
antisimmetrized spin states, denoted $S_{\uparrow\Uparrow}$,
$A_{\uparrow\Uparrow}$, etc, where thin and thick arrows correspond to
hole states with the angular momentum projection 1/2 and 3/2, respectively. 
The final three columns show the energies of the state for three radii
of a nanocrystal, as in the previous table.

\section*{References}

\providecommand{\newblock}{}


\end{document}